\documentclass[12pt]{article}
\usepackage[top=1in, bottom=1in, left=1in, right=1in]{geometry}
\usepackage{setspace}
\usepackage{graphicx} 
\usepackage{amsmath} 
\usepackage{caption} 
\usepackage{fancyhdr} 
\usepackage{titling} 
\usepackage{datetime} 

\usepackage[T1]{fontenc}
\DeclareSymbolFont{TOneChars}{T1}{\familydefault}{m}{it}
\SetSymbolFont{TOneChars}{bold}{T1}{\familydefault}{bx}{it} 
\DeclareMathSymbol{\mathdh}{\mathord}{TOneChars}{"F0}

\renewcommand{\maketitle}{
    \begin{titlepage}
        \centering
        \vspace*{2cm}
        {\bfseries \MakeUppercase{The interaction of gravitational waves with matter}\par}
        \vspace{2cm}
        { Nigel T. Bishop\textsuperscript{1}\textsuperscript{a}, Vishnu Kakkat\textsuperscript{2}\textsuperscript{b}, Amos S. Kubeka\textsuperscript{2}\textsuperscript{c}, Monos Naidoo\textsuperscript{1}\textsuperscript{d},\\ and Petrus J. van der Walt\textsuperscript{1}\textsuperscript{e} \\
        \vspace{.5cm}
        \textsuperscript{1}Department of Mathematics, Rhodes University, Grahamstown 6140,\\ South Africa\\
        \textsuperscript{2}Department of Mathematical Sciences, University of South Africa, P.O. Box 392,\\ Pretoria, South Africa} \\
        \vspace{1cm}
        Emails: \textsuperscript{a} n.bishop@ru.ac.za, \textsuperscript{b}kakkav@unisa.ac.za, \textsuperscript{c}kubekas@unisa.ac.za, \textsuperscript{d}monos.naidoo@gmail.com, \textsuperscript{e}peetvdw@worldonline.co.za\\
        \vspace{0.5cm}
        Corresponding author: kakkav@unisa.ac.za
        \vspace{2cm}
        \begin{abstract}
         
It is well-known that gravitational waves undergo no absorption or dissipation when traversing through a perfect fluid. However, in the presence of a viscous fluid, GWs transfer energy to the fluid medium. In this paper, we present a review of our recent series of results regarding the interaction between gravitational waves and surrounding matter. Additionally, we examine the impact of a viscous fluid shell on gravitational wave propagation, focusing particularly on GW damping and GW heating. Furthermore, we explore the significance of these effects in various astrophysical scenarios such as core-collapse Supernovae and primordial gravitational waves.
        \end{abstract}
        \vspace{1cm}
        \begin{center}
        \end{center}
        \vfill
    \end{titlepage}
}

\begin{document}

\maketitle

\section{Introduction}
Einstein predicted the existence of gravitational waves (GWs) in 1916~\cite{Einstein1916,Einstein1918}, but their physical properties and whether they carried energy remained unresolved for many years. In 1957, Feynman presented a thought experiment to the first American conference on general relativity at Chapel Hill, demonstrating that energy is transferred from gravitational waves to matter. Feynman considered a stick on which two massive particles could slide and GWs propagating transversely to the stick. The resultant movement of the particles along the stick would generate friction so releasing heat. In the next few years, the theory of GWs became fully developed, and a precise formula was obtained for the energy propagated by GWs~\cite{bondi1962gravitational}.

Feynman's argument shows that GWs would not transfer energy to dust or a perfect fluid because they are frictionless, but there would be an energy transfer to a viscous fluid. In 1966, Hawking~\cite{hawking1966perturbations} considered GW propagation in different cosmological models. It follows that
\begin{equation}
H(r+\Delta)=H(r)\exp\left(-\frac{8\pi G \eta \Delta}{c^3}\right)\,,
\label{e-H}
\end{equation}
where $H$ is the magnitude of the metric perturbation, $\eta$ is the fluid viscosity, and $\Delta$ is the distance propagated through the viscous fluid. In SI units, $G/c^3={\mathcal O}(10^{-36})$ and $\eta={\mathcal O}(1)$kg/m/s for everyday fluids. Although cosmological distances are large (1Gpc=${\mathcal O}(10^{25})$m), and there are extreme astrophysical conditions under which the effective viscosity can also be large (${\mathcal O}(10^{25})$kg/m/s), there are no known situations in which the damping represented by Eq.~\eqref{e-H} even approaches being significant.

Thus, over the last 50 years, it has been understood that GWs do interact with matter, but it has always been assumed that the effect is too small to be of any astrophysical or cosmological consequence. However, in a series of recent papers~\cite{bishop2020effect,Naidoo:2021,bishop2022effect, kakkat2024gravitational} we have shown that Eq.~\eqref{e-H} needs to be corrected when GWs interact with matter that is of order a wavelength away from the source, and this is astrophysically significant. There are feasible values of the parameters that lead to almost complete damping of GWs emitted by a core collapse supernova (CCSNe), and also of primordial GWs emitted during inflation. If an accretion disk were present around merging black holes, it would be heated sufficiently to produce an X-ray flash. Further, non-viscous matter such as dust can also modify emitted GWs causing a (frequency dependent) phase shift and echoes, leading to small but measurable modifications to the post-merger signal from a binary neutron star merger. There are further examples, discussed later in this paper.

This work uses the Bondi-Sachs~\cite{bondi1962gravitational,Sachs62} formalism for the Einstein equations; see also the reviews~\cite{Winicour2012,bishop2016extraction}. The method of separation of variables is used to construct eigensolutions for linearized perturbations of the metric~\cite{bishop2005linearized}; when the background is Minkowski spacetime, these solutions have a remarkably simple analytic form. Since the metric is known, it is straightforward to find the velocity field and hence the shear $\sigma_{ab}$ of the fluid flow. Then the rate of heating of the fluid is given by $-2\eta\sigma_{ab}\sigma^{ab}$ (where $\eta$ is the viscosity), which is balanced by an energy loss of the GWs passing through the fluid. It was found~\cite{bishop2022effect} that the GW damping effect is given by
\begin{equation}
\frac{dH}{dr}=-8\pi\eta H\frac{G}{c^3}\left(1+\frac{2}{r^2\nu^2}+\frac{9}{r^4\nu^4}+\frac{45}{r^6\nu^6}+\frac{315}{r^8\nu^8}\right)\,,
\label{e-dH}
\end{equation}
where $r$ is the distance form the source and $\nu/(2\pi)$ is the GW frequency. When $r\nu\gg 1$ the terms in brackets in Eq.~\eqref{e-dH} simplify to 1, and a simple calculation then leads to Eq.~\eqref{e-H}; but when $r\nu\ll 1$ the damping effect can be large. The metric perturbations have leading asymptotic order $1/r$ and such terms lead to the $1$ in the brackets in Eq.~\eqref{e-dH}. However, the metric perturbations also include other terms of order up to $1/r^3$, so leading to the additional terms in Eq.~\eqref{e-dH}.

This paper is structured as follows: Section \ref{sec_mattershell} introduces matter shell theory, GW damping, and GW heating. Section \ref{sec_astroapplica} delves into the astrophysical contexts where GW heating and GW damping is significant. Finally, the concluding section summarizes the key points.

\section{Matter shell around the GW source}
\label{sec_mattershell}


 To naturally determine the GW from different astrophysical sources we foliate the space-time by outgoing null hypersurfaces. These null hypersurfaces emanate from the timelike world tube (3-surface) $\Gamma$ which we construct by transporting a convex 2-surface along a timelike vector field. We uniquely describe the 2-surface by choosing angular coordinates ($x^{A},A=2,3$, e.g. in spherical polars $x^{2}=\theta,x^{3}=\phi$). We also describe $\Gamma$ by the coordinates $(u,x^{A})$ where $u$ is the time-coordinate along $\Gamma$. Furthermore, we label the null hypersufaces by the coordinates $x^{\alpha}=(u,r,x^{A})$, $\alpha=0,1,2,3$ where $x^{0}=u=const.$ and $x^{1}=r$ is the surface area coordinate.

Then the Bondi-Sachs metric in null coordinates takes the following form 
\begin{eqnarray}
ds^2=&-&\left[e^{2\beta}\left(1+\frac{W_c}{r}\right)-r^2h_{AB}U^{A}U^{B}\right]du^2-2e^{2\beta}dudr\nonumber\\
     &-&2r^2h_{AB}U^{B}dudx^{A}+r^{2}h_{AB}dx^{A}dx^{B},
\label{eq:nulla}
\end{eqnarray}
where $h^{AB}h_{BC}=\delta^{A}_{B}$ and $\det(h_{AB})=\det(q_{AB})$,
with $q_{AB}$ being a unit sphere metric (e.g. $q_{AB}dx^Adx^B=d\theta^2+\sin^2\theta d\phi^2$) and $h_{AB}$ the conformal (angular) metric. We introduce spin-weighted fields $U=U^{A}q_{A}$ and $J=q^{A}q^{B}h_{AB}/2$, where $q^A$ is a complex dyad, e.g. $q^A=(1,i/\sin\theta)$; see more details in \cite{newman1966note,gomez1997eth,bishop2016extraction}. The use of $J$ and $U$ leads to significant simplifications of the Einstein equations, and also enables the use of a simple separation of variables ansatz in Eqn.~\eqref{e-ansatz} below. Note that for the Schwarzschild spacetime, we have $J$ and $U$ being zero, and thus $J, U$ can be regarded as a measure of the deviation from spherical symmetry. 



In the following sections, we model the matter shell surrounding the GW source. Section \ref{thinshell} examines a thin, low-density shell, while Section \ref{viscoushell} focuses on the viscous shell around the GW source.
\subsection{Thin matter shell problem}
\label{thinshell}
\begin{figure}[ht]
\includegraphics[width=6cm]{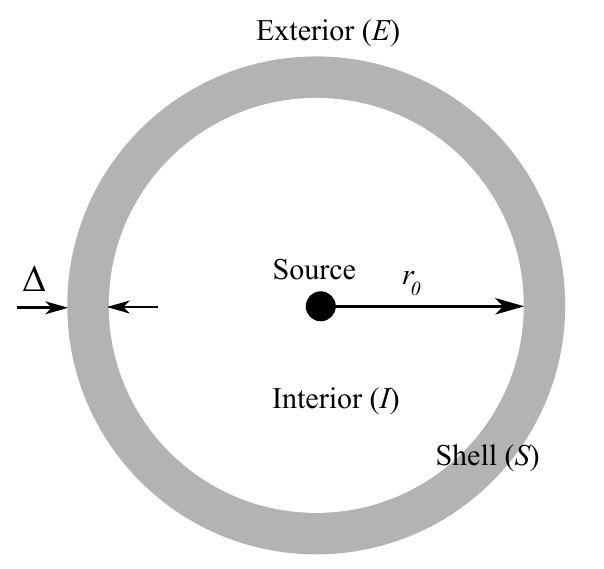}
\centering
\caption{A pictorial representation of a GW source surrounded by a spherical shell of mass $M_S$ located between $r = r_0$ and $r = r_0 + \Delta$ (where $r$ is the distance from the source).}
\label{fig1}
\end{figure}

We consider a static matter distribution as a thin, low-density shell surrounding a GW source, illustrated in Figure~\ref{fig1}. The shell is spherical at $r=const.$, with the density $\rho$ allowed to vary around the shell, making the problem generally not spherically symmetric. 
We make the ansatz of small quadrupolar perturbations about Minkowski spacetime with the metric quantities $(\beta,U,W,J)$ taking the form
\begin{align}
		\beta=&\Re(\beta^{[2,2]}(r)e^{i\nu u}){}_0Z_{2,2}\,,\;\;
		U=\Re(U^{[2,2]}(r)e^{i\nu u}){}_1Z_{2,2}\,,\nonumber \\
		W_c=&\Re(W_c^{[2,2]}(r)e^{i\nu u}){}_0Z_{2,2}\,,\;\;
		J=\Re(J^{[2,2]}(r)e^{i\nu u}){}_2Z_{2,2}\,,
		\label{e-ansatz}
	\end{align}
with the superfix ${}^{[2,2]}$ indicating a coefficient of ${}_s Z_{2,2}$. 	Note that the perturbations oscillate in time with frequency $\nu/(2\pi)$. The quantities ${}_s Z_{\ell,m}$ are spin-weighted spherical harmonic basis functions related to the usual ${}_s Y_{\ell,m}$ as specified in~\cite{bishop2005linearized,bishop2016extraction}. Notably, ${}_0 Z_{\ell,m}$ are real, allowing the description of the metric quantities $\beta$ and $W_c$ (which are real) without mode-mixing. However, for $s\ne 0$ ${}_s Z_{2,2}$ is generally complex. A general solution may be constructed by summing over the $(\ell,m)$ modes.

As shown in previous work~\cite{bishop2005linearized,bishop2020effect}, solving the vacuum Einstein equations under the condition of no incoming radiation leads to
\begin{align}
\beta^{[2,2]} ={}& b_0, \nonumber \\
    W_c^{[2,2]} ={}& 4i\nu b_0 - 2\nu^4 C_{40} - 2\nu^2 C_{30} + \frac{4i\nu C_{30} - 2b_0 }{r}+\frac{4i\nu^3 C_{40}}{r}+ \frac{12 \nu^2 C_{40}}{r^2} \nonumber \\
    &- \frac{12i\nu C_{40}}{r^3} - \frac{6C_{40}}{r^4}, \nonumber
\end{align}
\begin{align}
 U^{[2,2]} ={}& \frac{\sqrt{6}(-2i\nu b_0 + \nu^4 C_{40} + \nu^2 C_{30})}{3} + \frac{2\sqrt{6} b_0}{r}+ \frac{2\sqrt{6} C_{30}}{r^2} - \frac{4i\nu\sqrt{6} C_{40}}{r^4} \nonumber \\
    & - \frac{3\sqrt{6} C_{40}}{r^4}, \nonumber \\
    J^{[2,2]} ={}& \frac{2\sqrt{6}(2b_0 + i\nu^3 C_{40} + i\nu C_{30})}{3} + \frac{2\sqrt{6} C_{30}}{r} + \frac{2\sqrt{6} C_{40}}{r^3} ,
    \label{e-pert}
\end{align}
	with constants of integration $b_0, C_{30}, C_{40}$. The constant $C_{40}$ is physical and represents the magnitude of the GW source, while the constants $b_0$ and $C_{30}$ represent gauge freedoms. Denoting the news for the solution Eqn.~(\ref{e-pert}) by ${\mathcal N}_0$, and allowing for the conventions used here, we find ${\mathcal N}_{0}=-\sqrt{6}\nu^3 \Re(iC_{40}\exp(i\nu u))\,{}_2Z_{2,2}$. In the absence of matter shell, the gravitational wave strain, ${H}_{M0}=h_++ih_\times$, relates to the gravitational wave news ${\mathcal N}_{0}$ through ${H}_{M0}=2\int{\mathcal N}_{0}du$, where $h_+$ and $h_\times$ represent the polarization modes of the gravitational waves in the transverse-traceless (TT) gauge~\cite{naidoo2021modifications}. Therefore,
 \begin{equation}
   {H}_{M0}=\Re(\mathbf{H}_{M0}\exp(i\nu u))~{}_2 Z_{2,2}~~\textrm{with}~~\mathbf{H}_{M0}=-2\sqrt{6}\nu^2 C_{40}.
 \end{equation}

It is found that the dust shell modifies the gravitational wave both in magnitude and phase, without any energy being transferred to or from the dust \cite{bishop2020effect}. The introduction of a spherical shell around the GW source of mass $M_S$, radius $r_0$ and
thickness $\Delta$ modifies the wave strain to:
\begin{equation}\label{eqn_H}
 {H}=\Re\left(\mathbf{H}_{M0}\left(
1+\frac{2M_S}{r_0}+\frac{2iM_S}{r_0^2\nu}+\frac{i M_Se^{-2ir_0\nu}}{2r_0^2\nu}+{\mathcal O}\left(\frac{M_S\Delta}{r^2_0},\frac{M_S}{r_0^3\nu^2} \right)\right)\exp(i\nu u)\right)~{}_2 Z_{2,2}.
\end{equation} 
Notice that in the absence of the shell ($M_S=0$), $H=H_{M0}$ so that the GW is unmodified. Each term containing $M_S$ in Eqn.~\eqref{eqn_H} represents a correction to the GW due to the shell:
\begin{itemize}
    \item The first correction, ${2M_S}/{r_0}$, is part of the gravitational red-shift effect, the main consequence of which is a reduction in the frequency; this effect is well-known, and henceforth we will assume that GW waveforms to be considered have allowed for this effect.
    \item The second correction term, $iM_S/\pi r_o f$, is out of phase with the leading terms
${1 + 2M_S /r_0}$ and hence represents a phase shift of the GW. This term does not affect the energy of the GW.
\item The presence of $e^{-4\pi ir_0f}$
in the third correction term describes a change in the
magnitude of $H$ and is due to an incoming wave modifying the
geometry near the source and thus the inspiral rate.
\end{itemize}

\subsection{Viscous shell}
\label{viscoushell}
We have discussed, previously, that a dust shell surrounding a gravitational wave (GW) event modifies a GW wave in both magnitude and phase. In this section, we discuss the effect of a viscous shell on GW propagation. Now turning our focus towards a viscous shell enveloping a gravitating source at a finite distance $r$, characterized by coefficients of shear viscosity $\eta$ and bulk viscosity $\zeta$. It has been elucidated in \cite{bishop2022effect} that when GWs traverse through this viscous shell, the shear viscosity $\eta$ influences the propagation of GWs while the bulk viscosity does not. The rate of change of GW energy loss due to viscosity is given by
\begin{equation}
    \left<\dot{E}_\eta\right> = -12\eta  C_{40}\nu^6 \delta r
		\left(1+\frac{2}{r^2\nu^2}+\frac{9}{r^4\nu^4}
		+\frac{45}{r^6\nu^6}+\frac{315}{r^8\nu^8}
		\right),
\end{equation}
where $\delta r$ represents the thickness of the shell and $\left<f\right>$ represents the average of $f$ over the wave period.
The rate of energy loss $\left<\dot{E}_\eta\right>$ calculated using the expression $-2\eta \sigma^{ab}\sigma_{ab}$, $\sigma_{ab}$ is the shear tensor.  The negative sign shows the flow of energy from the GWs to the viscous shell. Furthermore, the energy dissipated inside the viscous shell can be expressed in terms of the GW energy as
\begin{equation}
		\left<\dot{E}_\eta\right> = -16\pi\eta \delta r \left<\dot{E}_{GW}\right>
		\left(1+\frac{2}{r^2\nu^2}+\frac{9}{r^4\nu^4}
		+\frac{45}{r^6\nu^6}+\frac{315}{r^8\nu^8}
		\right).
  \end{equation}
Due to the dissipation of energy caused by  GWs, two phenomena may occur. The first is the damping of GWs, and followed by the heating of the viscous shell. The subsequent section delves into the concepts of GW damping and GW heating.
\subsubsection{GW damping and GW heating}
\label{sec_viscousshell}

To see the damping of the GWs inside the viscous shell, let $H$ denote the magnitude of the GWs rescaled to account for a $1/r$ falloff, and let $r_o$ and $r_i$ represent the outer and inner radii of the shell, respectively. As $r_o$ tends to infinity, the magnitude of the GWs can be expressed as~\cite{bishop2022effect}:
\begin{equation}
		{H}(r_o)={H}(r_i)\exp\left(-\frac{45\eta\lambda^8}{32r_i^7\pi^7}\right)\,,
		\label{h2}
	\end{equation}
 where $\lambda$ is the wavelength of the GWs.
 It is notable that the magnitude of the GWs decays exponentially, with the damping effect only marginally reduced for a shell of finite thickness. Furthermore, there exist astrophysical scenarios where the damping is nearly complete. See more detailed analysis on the GW damping in various astrophysical cases in \cite{bishop2022effect}.

An intriguing question stemming from this analysis is whether the energy transferred from the GWs to the shell leads to its heating.  Our recent findings in this regard, as detailed in our paper \cite{kakkat2024gravitational}, outline a process through which the viscous shell can indeed be heated.  It is shown previously that the energy input to the shell is a function of the angular coordinates. However, for the applications to be considered here in the subsequent sections, the angular dependence of the GWs is uncertain, so we treat the heating effect as uniform across the shell. Then the $Y_{0,0}$ terms in the formulas in~\cite{kakkat2024gravitational} are retained, but those terms with higher order spherical harmonics ($Y_{2,0},Y_{4,0}$) are discarded. Therefore Eq.~(10) of~\cite{kakkat2024gravitational} simplifies to
\begin{equation}\label{eqn:eshell}
    \frac{\partial_u E_{shell}}{\Delta V}=\frac{\sqrt{\pi}}{6}\nu^2\eta \partial_u E_{GW}D_0\,,
\end{equation}
where $\partial_u E_{GW}$ is the power output of GWs, and where
\begin{equation}
D_0= \frac{12({\nu}^8r^8+2{\nu}^6r^6+9{\nu}^4r^4+45{\nu}^2r^2+315)}{\sqrt{\pi}{\nu}^{10}r^{10}}.
\label{e-TA-D}
\end{equation}
Then the temperature increase in the shell is
\begin{equation}
T-T_0=\frac{\sqrt{\pi} G\eta}{6c^5 C\rho}\nu^2  \Delta E_{GW}D_0\,,
\label{eqn:diffT}
\end{equation}
where conversion to SI units has been made, with $G$ the gravitational constant and $c$ the speed of light, and where $\rho$ is the density and $C$ is the specific heat capacity (in J/${}^\circ$K/kg). Note that in the formula for $D_0$ in Eq.~\eqref{e-TA-D}, $r\nu\rightarrow r \nu/c^2$.

Furthermore, in certain circumstances, the temperature within the shell can rise to temperatures on the order of $10^6~K$. The subsequent section will elucidate the impact of damping and GW heating across various astrophysical scenarios where it holds significance.

\section{Astrophysical and cosmological applications}
\label{sec_astroapplica}
\subsection{Core Collapse Supernovae}

Core Collapse Supernovae (CCSNe) are expected to produce GWs detectable by the current generation of interferometers or those on the horizon.  
The anticipated GW signal from CCSNe, is normally described by four phases. Initially there is the  convection signal. This is followed by a quiescent phase. The third phase is driven by the standing-accretion-shock instability (SASI) and is also referred to as the neutrino convection phase. Finally, there is the  explosion phase.
To date, detection of supernovae (SN) has only been through  electromagnetic waves~\cite{Pennypacker:1988xt,Fabian:1995ik,Montes:1997sq} and neutrinos~\cite{Kamiokande-II:1987idp}. 
However, as photons originate at the outer edge of a star, the information provided so far in SN is limited. The physics of the interior regions is still poorly understood. In a SN explosion, where all the four fundamental forces of nature are involved, the aspherical motion of the inner regions will produce GWs. Detection of these GWs will provide a wealth of information on the the central engines and inner regions of CCSNe.  
Discussions on the theory, nature, and explosion mechanisms of CCSNe ~\cite{Janka:2006fh,Smartt:2009zr,Bethe:1990mw,Janka:2012wk}
are limited by the uncertainties of the interior of the CCSNe progenitors, making numerical modelling  of these progenitors ~\cite{Muller:2020ard,Abdikamalov:2020jzn} also difficult. 
CCSNe are expected for stars with masses larger than 8$M_\odot$
but less than 100$M_\odot$.
For such stars, evolution normally proceeds through several stages of core burning. Once nuclear fusion halts, which is typically, at the stage of reaching an iron core, after which no nuclear burning can take place, then there are no further burning processes to balance the gravitational attraction, leading to the onset of core collapse. The core breaks into two during the collapse. The outer core collapses supersonically. However, the inner core of $0.4M_{\odot}$ to $0.6M_{\odot}$ remains in sonic contact and collapses homologously, reaching supranuclear densities of $\sim 2 \times 10^{14}$gm/cm$^{3}$. This causes the nuclear matter to stiffen, resulting in a bounce of the inner core and launching the resulting shock wave into the collapsing outer core. This shock, in turn, loses energy to dissociation of iron nuclei, stalling at ${\sim} 150\,\mathrm{km}$ within ${\sim} 10\,\mathrm{ms}$ after formation.

\subsubsection{CCSNe as a dust shell}
For a dust shell surrounding a GW emission, modifications to the GW occur in both magnitude and phase~\cite{Bishop:2019eff}. 
In~\cite{Naidoo:2021}, this analysis was extended to the GW event associated with a CCSNe, treated the PNS as a dust shell surrounding a GW emission and, again, it was shown that modifications to the GW occur in both magnitude and phase.  The results were shown to be potentially astrophysically relevant with the modification measurable for certain parameters. The magnitude was shown to be proportional to $\lambda/r_i$, with $\lambda$ the GW wavelength and $r_i$ the inner radius of the matter shell.
It was also shown that a burst of GWs can, in principle, lead to echoes~\cite{Naidoo:2021}, although the time difference between the echo and the actual wave and the relative amplitude of the echo to the main signal would mean the echo would be buried in the main signal and hence not discernible.

\subsubsection{Viscosity in the core collapse environment}
Not much attention has been in the literature to viscosity in the core collapse environment. Potential mechanisms for viscosity have been suggested in~\cite{Thompson:2004if} to include neutrino viscosity, turbulent viscosity caused by the magnetorotational instability (MRI), and turbulent viscosity by entropy and composition-gradient-driven convection. Of these, the MRI was found to be the most effective. The MRI dominates the neutrino viscosity by 2 to 3 orders of
magnitude. The authors~\cite{Thompson:2004if} also found that, within the PNS, the  MRI will operate and dominate the viscosity even for the slowest rotators they considered.

In \cite{bishop2022effect}, we investigated, again using the Bondi-Sachs form of the Einstein equations linearized about Minkoski, the effect of viscosity on GW propagation. A general expression for the damping effect was found. With $r_i,r_o$, the inner and outer radii of the shell respectively, a novel result was found for a shell of finite thickness $(r_o-r_i)$ where $r_i$ is much smaller than the wavelength $\lambda$ of the GWs. To our knowledge, viscous damping of GWs with $r_i\ll\lambda$ had not been studied previously.

The paper then looked at the astrophysical application of CCSNe. A number of CCSNe models with different parameters were proposed. Detectable frequencies of GWs expected from CCSNe, range from 100 to 1300 Hz. Estimates for the radius of the core of the PNS range from 10 to 30 km. The GW energy released from a CCSNe is expected to be approximately $10^{43} -10^{46}$ erg, with the density at core bounce expected to be approximately $3.2\times 10^{14}$~{kg/m}${}^3$  and the dynamic viscosity ranges from $10^{23}$ to $10^{25}$\,kg/m/s. Significant viscous damping of GWs was predicted for many cases ranging from a discernible 10\% to complete damping. GW generation in CCSNe involves a number of different physical processes, each with GW output at a different frequency. It may be that a GW observation of a CCSNe event would see only the higher frequencies, with the lower ones completely damped out.

\subsubsection{GW Heating in CCSNe}
Following on the work in \cite{kakkat2024gravitational} we then consider whether there is any heating effect on the PNS owing to the viscous damping (in preparation). 
We find
\begin{equation}
38\,\mbox{K}\le (T-T_0)\le 7.0\times 10^{20}\,\mbox{K}\,,
\label{e-CSSNe-dT}
\end{equation}
so that, depending on the actual values of physical parameters, the significance of the effect varies from very high to none.
For realistic scenarios, it seems that GW heating of the PNS is a likely scenario. One of the consequences of such heating is that, in the case of a shock wave stalling in a post-bounce phase, GW heating may produce enough energy to resuscitate the shock wave, hence ensuring the PNS results in a CCSNe where it otherwise may not have. 

\subsection{Primordial gravitational waves}

Through electromagnetic radiation, the Universe is only directly observable back in time to the surface of last scattering via the cosmic microwave background (CMB). Although earlier processes can be inferred by analysing subtle structures in the CMB, direct observation can only achieved by other means. Due to the expectation that GWs are only weakly interacting with matter of all kinds, it is anticipated that primordial gravitational waves (pGWs) will provide this essential window into the evolution of the very early Universe. Understanding the nature of these early GWs, is further of interest to the fundamental knowledge of processes in high-energy physics that are well beyond the capabilities of Earth-based experiments. In order to enhance the understanding of how the detection of pGWs should be anticipated, we consider the effects of viscosity on pGWs in this section.

	In early works, such as that of Hawking in 1966 \cite{hawking1966perturbations}, it was already shown that shear viscosity, $\eta$, will affect pGWs through damping (Eq. \eqref{e-H}). These results along with the work of Eckart in the 1940s \cite{Eckart1940-od} were consolidated and further developed into what can be considered the seminal work on primordial dissipative processes by Weinberg in 1971 \cite{Weinberg1971}. Since then, there has been significant progress in understanding the details of early epochs and their physical properties but the work of Weinberg is still relevant as baseline formulae for relativistic viscous effects. In \cite{bishop2022effect}, we described representative early Universe scenarios to illustrate the viscous shell effects on pGW in the extreme conditions expected in the epoch after cosmic inflation ends, which we summarise in the following section. While recognising that pGWs generated during inflation are expected to be fundamentally stochastic (see \cite{Caprini2018}), for our purposes here, we will consider a discrete source of pGWs and reserve extending our work to better model the stochastic nature of pGWs for future research.
	
	\subsubsection{Cosmological scenario}

The results in the previous sections, where the effects of viscous damping are dependent on shell radii, can be translated to scenarios in cosmology by relating distances to durations. In particular, by considering the relationship between the temperature $T$ in early epochs to the cosmological scale factor $a(t)$ and the Hubble rate $\mathcal{H}$ (denoted here by $\mathcal{H}$ instead of the usual $H$ to avoid confusion with the metric perturbation, defined earlier), Eq. \eqref{e-dH} can be solved in terms of an initial and end time of a cosmological epoch.

An illustration follows by considering a hypothetical scenario where GWs propagate from a single source subsequent to the epoch of cosmic inflation in a radiation-dominated Einstein-de Sitter (EdS) universe with $a(t) \propto t^{1/2}$. At the end of inflation at $t=10^{-32}$s and $T=10^{27}$ to $10^{28}$K. According to \cite[Section 28.1]{Misner1973-mq}, $T$ is proportional to $1/a(t)$. Thus, at the end of inflation, we take $a=10^{-27}$. Using the start of inflation as $10^{-36}$s with the Hubble rate not changing, due to exponential expansion, the inverse Hubble rate is taken as $1/\mathcal{H} (=a/\dot{a})= 10^{-36}$s, which leads to the horizon scale as $c/\mathcal{H} = 3 \times 10^{-28}$m.

In terms of GW properties, we consider the ranges of detection rather than the physical processes generating the waves. For this, ground-based GW detectors operate, approximately, in the frequency range $10$ to $10^3$Hz; pulsar timing arrays and the planned satellite system LISA will extend the lower limit to about $10^{-9}$Hz. Thus searches for pGWs will cover wavelengths in the range $3 \times 10^5$m to $3 \times 10^{17}$m. Due to cosmological redshift, wavelengths scale as $a(t)$, which corresponds to a wavelength range at the end of inflation of $3 \times 10^{-22}$ to $3 \times 10^{-10}$m.

An estimate for viscosity, from Weinberg 1971 \cite{Weinberg1971}, using Eqns (3.20--3.22), rewritten with $c\ne 1$, we get \footnote{Even though there are more recent results of $\eta$, we take the value derived from Weinberg as a baseline for an extreme viscosity under ultra-high energy conditions to illustrate the damping effect.}
	\begin{equation}
		\eta^2=\frac{aT^4c}{60\pi G} .
	\end{equation}
	When $T=10^{27}$K,
	\begin{equation}
		\eta=3.68 \times 10^{58} \, \mathrm{kg/m/s} \,.
		\label{e-eta-W}
	\end{equation}
	This value of $\eta$ is very large, and even when multiplied by $G/c^3$ to convert to geometric units, we get a value of $9.09 \times 10^{22}$m${}^{-1}$.

If we consider a thin shell scenario where  $r_i,r_o$ are much larger than the wavelength $\lambda$, solving Eqn.~\eqref{e-dH} for $r\nu \gg 1$ and rewriting in terms of $t$, we have
	\begin{equation*}
		\frac{d{H}}{dr}=-8\pi\eta {H} \rightarrow \frac{d{H}}{dt}=-8\pi\eta c {H}
	\end{equation*}
in geometric units.
	Now, $\eta$ behaves as $T^2$, i.e. as $1/a^2$ and further, we are in the radiation-dominated era with $a(t) \propto t^{1/2}$. Thus $\eta$ behaves as $1/t$ and we have
	$dH/dt = - 8 \pi \eta_i c H t_i/t$ with $\eta_i=9.09 \times 10^{22}$ and $t_i=10^{-32}$.
	Let $A=8 \pi \eta_i c t_i =6.9$, then integrating we get
	\begin{align}
		{H}_o=H_i \left(\frac{t_i}{t_o} \right)^{A} .
	\end{align}
For example, suppose that
\begin{equation}
t_o=2t_i\;\;\mbox{then}\;\; {H}_o=0.009\,{H}_i\,.
\end{equation}
This example shows that if $\eta$ is as large as the value in Eqn.~(\ref{e-eta-W}) then the damping effect is so rapid that it may occur before the quark epoch is reached. Repeating the calculation for the upper estimate of the viscosity of a quark-gluon plasma, i.e., $\eta \le 5 \times 10^{11}$ kg/m/s (see \cite{Schafer2009}), further damping during the quark-gluon epoch appears to be inconsequential.

Alternatively, considering that the wavelength is in the range of $3 \times 10^{-22}$ to $3 \times 10^{-10}\,$m while the horizon scale is $c/\mathcal{H} = 3 \times 10^{-27}$\,m, the case for $r_i,r_o\ll\lambda$ can be justified and Eqn.~(\ref{h2}) becomes applicable. Using the same formulation, significant damping can then occur for even small values of $\eta_i$. As opposed to the first scenario where the value of $\eta$ dominates damping, this effect is present in geometries where the $(\lambda/r_i)$ ratio is large, which amplifies the value of the negative exponent in Eqn.~(\ref{h2}).

As an example, suppose that  $r_i=3 \times 10^{-28}$\,m (a pGW source the size of the horizon scale) with $\lambda=3 \times 10^{-22}$\,m (the lower wavelength limit), then ${H}_o=0.5 {H}_i$ for $\eta=2.0\times 10^{25}$\,kg/m/s. This value is much smaller than that of Eqn.~(\ref{e-eta-W}), and is of the same order as values considered for CCSNe. The timescale of the damping is $10^{-36}$\,s, which is much shorter than the timescale of inflation.
	
\section{Conclusion}
Recent years have seen a surge of interest in GW studies, largely due to the regular direct detections. As GWs travel from their source, they interact with matter in various ways, with some of these interactions discussed in~\cite{hawking1966perturbations,esposito1971absorption,marklund2000radio,brodin2001photon,cuesta2002gravitational}. Despite these interactions, GWs typically remain unaffected by matter, enabling them to traverse cosmological distances without significant attenuation. It's established that GWs experience no absorption or dissipation when passing through a perfect fluid, while energy is transferred from GWs to a viscous fluid~\cite{hawking1966perturbations}.

This essay delves into the astrophysical contexts where GW damping and GW heating are highly relevant. Specifically, we examine the effects of a matter shell enveloping the GW source. Initially, we analyze a low-density dust shell, providing the modified GW strain equation (Eqn.~\ref{eqn_H}). Expanding our discourse, we explore the implications of a viscous shell encircling the GW source. Energy dissipation occurs within this shell as GWs traverse it, leading to GW damping and heating of the shell.

Subsequently, we delve into astrophysical scenarios where GW damping and GW heating play crucial roles. Core-collapse supernovae, anticipated as the next frontier of GW direct detections, and primordial gravitational waves are among the key focuses. Ongoing efforts are directed toward further research in these areas.

	
\section*{Acknowledgements}
VK and ASK express their sincere gratitude to Unisa for the Postdoctoral grant and their generous support.


\pagestyle{empty}

\end{document}